\documentclass[apj]{emulateapj}

\newcommand{\cubecm}{\ifmmode{~{\rm cm^{-3}}}\else{~cm$^{-3}$}\fi}
\newcommand{\kms}{\ifmmode{~{\rm km~s^{-1}}}\else{~km s$^{-1}$}\fi}
\newcommand{\lsim}{\lower0.3em\hbox{$\,\buildrel <\over\sim\,$}}
\newcommand{\gsim}{\lower0.3em\hbox{$\,\buildrel >\over\sim\,$}}

\newcommand{\enzo}{{\sl Enzo}}
\newcommand{\Ms}{M_\odot}

\newcommand{\hh}{H$_2$}
\newcommand{\Ol}{$\Omega_\Lambda$}
\newcommand{\Om}{$\Omega_M$}
\newcommand{\Ob}{$\Omega_b$}
\newcommand{\tcool}{$t_{\rm{cool}}$}
\newcommand{\rcool}{$r_{\rm{cool}}$}

\newcommand{\tdyn}{$t_{\rm{dyn}}$}

\newcommand{\tH}{$t_{\rm{H}}$}
\newcommand{\tvir}{$T_{\rm{vir}}$}
\newcommand{\lya}{Ly$\alpha$}
\newcommand{\rvir}{\ifmmode{{\rm r_{vir}}}\else{r$_{\rm{vir}}$}\fi}
\newcommand{\mvir}{\ifmmode{{\rm M_{vir}}}\else{M$_{\rm{vir}}$}\fi}

\shorttitle{RESOLVING THE FORMATION OF PROTOGALAXIES II}
\shortauthors{WISE, TURK, \& ABEL}

\begin{document}

\title{Resolving the Formation of Protogalaxies. II. Central
  Gravitational Collapse}

\author{John H. Wise\altaffilmark{1,2}, 
  Matthew J. Turk\altaffilmark{1}, and 
  Tom Abel\altaffilmark{1}}

\altaffiltext{1}{Kavli Institute for Particle Astrophysics and
  Cosmology, Stanford University, Menlo Park, CA 94025}
\altaffiltext{2}{Current address: Laboratory for Observational
  Cosmology, NASA Goddard Space Flight Center, Greenbelt, MD 21114}
\email{john.h.wise@nasa.gov}

\begin{abstract}
  Numerous cosmological hydrodynamic studies have addressed the
  formation of galaxies.  Here we choose to study the first stages of
  galaxy formation, including non-equilibrium atomic primordial gas
  cooling, gravity and hydrodynamics.  Using initial conditions
  appropriate for the concordance cosmological model of structure
  formation, we perform two adaptive mesh refinement simulations of
  $\sim$$10^8 \Ms$ galaxies at high redshift.  The calculations
  resolve the Jeans length at all times with more than 16 cells and
  capture over 14 orders of magnitude in length scales.  In both
  cases, the dense, $10^5$ solar mass, one parsec central regions are
  found to contract rapidly and have turbulent Mach numbers up to 4.
  Despite the ever decreasing Jeans length of the isothermal gas, we
  only find one site of fragmentation during the collapse.  However,
  rotational secular bar instabilities transport angular momentum
  outwards in the central parsec as the gas continues to collapse and
  lead to multiple nested unstable fragments with decreasing masses
  down to sub-Jupiter mass scales.  Although these numerical
  experiments neglect star formation and feedback, they clearly
  highlight the physics of turbulence in gravitationally collapsing
  gas.  The angular momentum segregation seen in our calculations
  plays an important role in theories that form supermassive black
  holes from gaseous collapse.
\end{abstract}


%
%

\begin{deluxetable*}{ccccccc}
\tabletypesize{}
\tablewidth{0pc}
\tablecaption{Simulation Parameters\label{tab:params}}

\tablehead{\colhead{Name} & \colhead{l} & \colhead{N$_{part}$} &
  \colhead{N$_{grid}$} & \colhead{N$_{cell}$} & \colhead{L$_{max}$} &
  \colhead{$\Delta$x} \\
  \colhead{} & \colhead{[Mpc]} & \colhead{} & \colhead{} & \colhead{}
  & \colhead{} & \colhead{[R$_\odot$]}
}
\startdata
A & 1.0 & 2.22 $\times$ 10$^7$ & 44712 & 1.23 $\times$ 10$^8$
(498$^3$) & 41 & $9.3 \times 10^{-3}$ \\
B & 1.5 & 1.26 $\times$ 10$^7$ & 22179 & 7.40 $\times$ 10$^7$
(420$^3$) & 41 & $1.4 \times 10^{-2}$

\enddata
\tablecomments{Col. (1): Simulation name. Col. (2): Comoving size of
  the simulation. Col. (3): Number of dark matter particles. Col. (4):
  Number of AMR grids. Col. (5): Maximum number of unique grid
  cells. Col. (6): Maximum level of refinement reached in the
  simulation. Col. (7): Resolution at the maximum refinement level.}
\end{deluxetable*}

%
%

\begin{deluxetable*}{ccccccc}
\tabletypesize{}
\tablewidth{0pc}
\tablecaption{Halos of interest\tablenotemark{a}\label{tab:runs}}

\tablehead{\colhead{l} & \colhead{z} & \colhead{M$_{tot}$} &
\colhead{$\sigma$} &\colhead{$\rho_c$} & \colhead{T$_c$} &
\colhead{M$_{BE}$} \\
\colhead{[Mpc]} & \colhead{} & \colhead{[$\Ms$]}  & \colhead{} &
\colhead{[cm$^{-3}$]} &\colhead{[K]} & \colhead{[$\Ms$]}}

\startdata

1.0 & 15.87 & 3.47 $\times$ 10$^7$ & 2.45 & 5.84 $\times$ 10$^{21}$ & 
8190 & 4.74 $\times$ 10$^5$ \\

1.5 & 16.80 & 3.50 $\times$ 10$^7$ & 2.59 & 7.58 $\times$ 10$^{21}$ & 
8270 & 1.01 $\times$ 10$^5$

\enddata
\tablenotetext{a}{The subscript ``c'' denotes central quantities.}
\tablecomments{Col. (1): Box size of the simulation. Col. (2): Final
  redshift of simulation. Col. (3): Total halo mass. Col. (4):
  $\sigma$ of the total mass compared to matter
  fluctuations. Col. (5): Central halo density. Col. (6): Central gas
  temperature. Col. (7): Gravitationally unstable central mass.}
\end{deluxetable*}

\section{MOTIVATION \& PREVIOUS WORK}

Since the first investigations of galaxy interactions
\citep{Holmberg41} using light bulbs, the use of numerical simulations
in galaxy formation has developed dramatically.  Not only gravity but
also hydrodynamics and cooling are standard ingredients in the
sophisticated computer models studying galaxy formation and
interactions.  In hierarchical structure formation, dark matter (DM)
halos merge to form larger halos while the gas infalls into these
potential wells \citep{Peebles68, White78}.  \citeauthor{White78}
provided the basis for modern galaxy formation, in which small
galaxies form early and continuously merge into larger systems.

As more high redshift galaxies were observed in the following 10
years, \citet{White91} refined the theory to address the observed
characteristics in these galaxies.  In their model, the halo
accumulates mass until the gas cools faster than a Hubble time, \tH,
which usually occurs when atomic hydrogen line, specifically \lya,
cooling is efficient.  This happens when the halo has \tvir~$>$~10$^4$
K, where the cooling function sharply rises by several orders of
magnitude because the number of free electrons able to excite hydrogen
greatly increases at this temperature \citep{Spitzer78}.  One can
define a cooling radius, \rcool, in which the interior material is
able to cool within a Hubble time.  Once the halo reaches this first
milestone, \rcool~ increases through additional accretion and cooling.
A rapid baryonic collapse ensues when \tcool~$\lsim$~\tdyn~
\citep{Rees77}.  The material accelerates towards the center, and its
density quickly increases.  In the model discussed in White \& Frenk,
this collapse will halt when one of the following circumstances
occurs.  First, angular momentum can prevent the gas from collapsing
further, and the system becomes rotationally supported.  Afterwards,
this disc fragments and star formation follows.  Alternatively, star
formation does not necessarily develop in a disc component, but the
energy released by stars during their main sequence and associated
supernovae (SNe) terminates the collapse.

These concepts have been applied also to the earliest galaxies in the
universe \citep{Mo98, Oh02, Begelman06, Lodato06}.  Many studies
\citep[e.g.][]{Ostriker96, Haiman97b, Cen03, Somerville03, Wise05}
demonstrated that OB-stars within protogalaxies at $z > 6$ can produce
the majority of photons required for reionization.  These
protogalaxies contain an ample gas reservoir for widespread star
formation, and the accompanying radiation propagates into and ionizes
the surrounding neutral intergalactic medium.  Several high redshift
starburst galaxies have been observed that host ubiquitous star
formation at $z > 6$ \citep{Stanway03, Mobasher05, Bouwens06}.
Additionally, supermassive black holes (SMBH) more massive than 10$^8
\Ms$ are present at these redshifts \citep[e.g.][]{Becker01, Fan02,
  Fan06}.  Finally, a reionization signature in the polarization of
the cosmic microwave background (CMB) at z $\sim$ 10 \citep{Page07}
further supports and constrains stellar and SMBH activity at high
redshifts.

The distinction between SMBH formation and a starburst galaxy should
depend on the initial ingredients (i.e. seed BHs, metallicity, merger
histories) of the host halo, but the evolution of various initial
states is debatable.  It is essential to study the hydrodynamics of
high redshift halo collapses because the initial luminous object(s)
that emerges will dynamically and thermally alter its surroundings.
For example, as the object emits ultraviolet radiation, the nearby gas
heats and thus the characteristic Jeans mass increases, which may
inhibit the accretion of new gas for future star formation
\citep{Efstathiou92, Thoul96}.

The following work will attempt to clarify early galaxy formation by
focusing on protogalactic (\tvir~$>10^4$ K) halos and following their
initial gaseous collapse.  \citet[][hereafter Paper I]{Wise07a} studied
the virialization of protogalactic halos and the virial generation of
supersonic turbulence.  In this paper, we address the gas dynamics of
the continued, turbulent collapse of a halo and study the evolution
and characteristics of the central regions.  In later studies, we will
introduce the effects from primordial star formation and feedback and
\hh~cooling.  The progressive introduction of new processes is
essential to understand the relevance of each mechanism.  We argue
that our results may be relevant for scenarios that envisage SMBH
formation from gaseous collapses.

\citet{Loeb94} and \citet{Bromm03} conducted smoothed particle
hydrodynamics (SPH) simulations that focused on the collapse of
idealized, isolated protogalactic halos.  The former group concluded
that a central $10^6 \Ms$ SMBH must exist to stabilize the thin
gaseous disc that forms in their calculations.  \citeauthor{Bromm03}
considered cases with and without \hh~chemistry and a background UV
radiation field.  They observed the formation of a dense object with a
mass $M \sim 10^6 \Ms$, or $\gsim 10\%$ of the baryonic matter, in
simulations with no or strongly suppressed \hh~formation.  These
calculations without metal cooling and stellar feedback are useful to
explore the hydrodynamics of the collapse under simplified conditions.
\citet{Spaans06} analytically studied the collapse of 10$^4$ K halos
with an atomic equation of state.  They find that $\sim$0.1\% of the
baryonic mass results in a pre-galactic BH with a mass $\sim$$10^5
\Ms$.  \citet{Lodato06} also found that $\sim$5\% of the gas mass in
$M = 10^7 \Ms$ halos at $z \sim 10$ becomes unstable in a gaseous disc
and forms a SMBH.  Recently, \citet{Clark07} studied the effects of
metal and dust cooling on the fragmentation of a collapsing
protogalactic core with varying metallicities ($Z = 0, 10^{-6},
10^{-5} Z_\odot$) and found the gas fragmenting ten times as much in
the $10^{-5} Z_\odot$ case than the primordial case.  In addition, the
fragments in the primordial case are biased toward larger masses.

A runaway gaseous collapse requires angular momentum transport so
material can inflow to small scales and form a central object.  The
stability of rotating gaseous clouds have been subject of much
interest over the last four decades and was thoroughly detailed by
the work of \citet[][hereafter EFE]{Chandra69}.  In the 1960's and
1970's, studies utilizing virial tensor techniques
\citep[EFE;][]{Lebovitz67, Ostriker69, Ostriker73a}, variational
techniques \citep{LyndenBell67, Bardeen77}, and N-body simulations
\citep{Ostriker73b} all focused on criteria in which a stellar or
gaseous system becomes secularly or dynamically unstable.  The first
instability encountered is an $m = 2$ bar-like instability that is
conducive for angular momentum transport in order to form a dense,
central object.  \citet{Begelman06} investigated the conditions where
a gaseous disc in a pre-galactic halo would become rotationally
unstable to bar formation \citep[see][]{Christodoulou95a,
  Christodoulou95b}.  They adapt the ``bars within bars'' scenario
\citep{Shlosman89, Shlosman90}, which was originally formulated to
drive SMBH accretion from a gaseous bar that forms within a stellar
galactic bar, to the scenario of pre-galactic BH formation.  Here a
cascade of bars form and transport angular momentum outwards, and the
system can collapse to small scales to form a quasistar with runaway
neutrino cooling, resulting in a central SMBH.  The simulations
detailed below show how many central bar-like instabilities form.


In \S2, we describe our simulations and their cosmological context.
In the following section, we present our analysis of the halo collapse
simulations and investigate the structural and hydrodynamical
evolution, the initial halo collapse, rotational instabilities, and
the importance of turbulence.  In \S4, we discuss the relevance of
angular momentum transport and rotational instabilities in early
galaxy and SMBH formation.  There we also examine the applicability
and limitations of our results and desired improvements for future
simulations.  Finally we conclude in the last section.

%
%

\begin{figure*}
  \resizebox{\textwidth}{!}{\rotatebox{0}{\includegraphics*{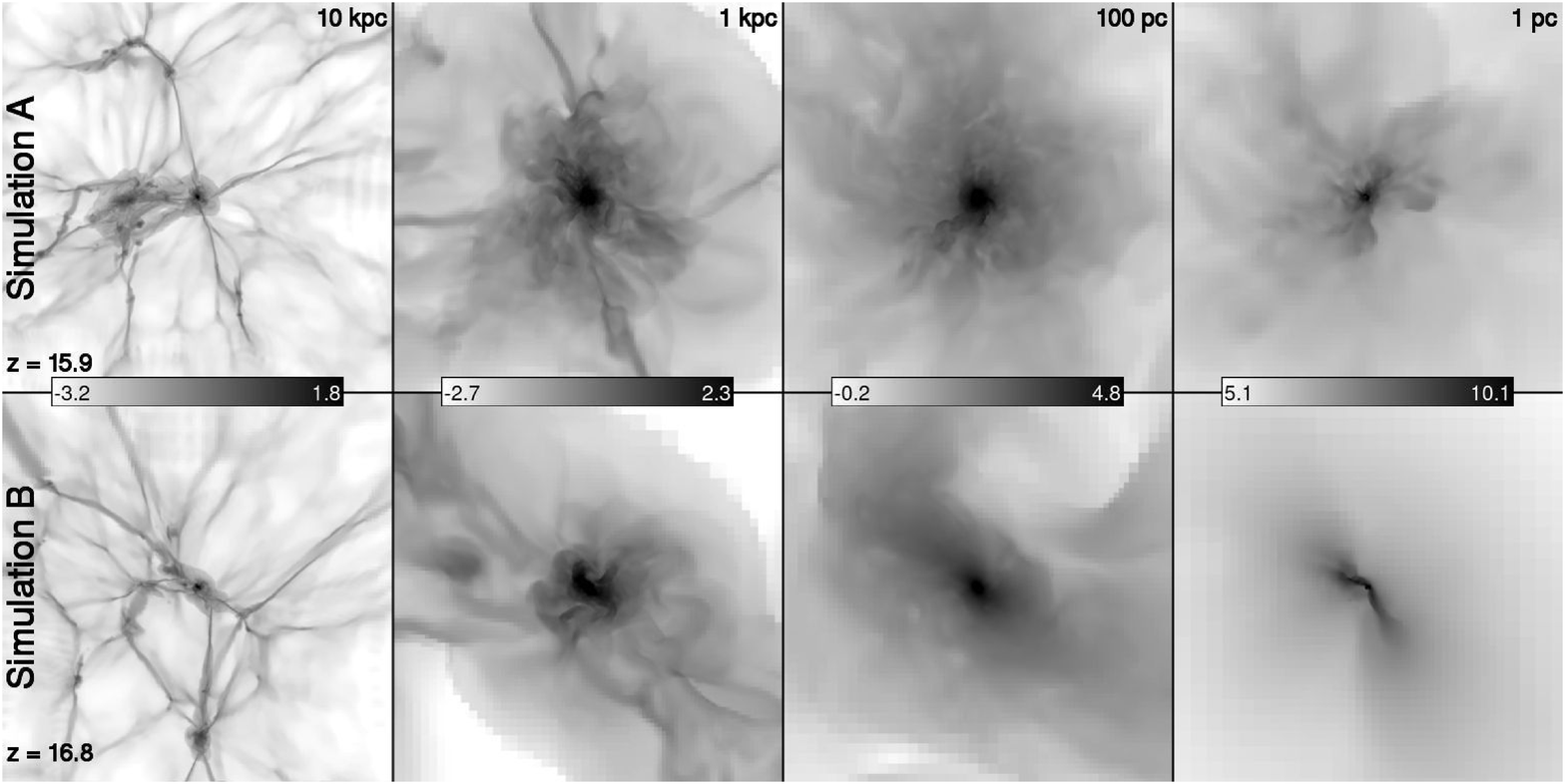}}}
  \resizebox{\textwidth}{!}{\rotatebox{0}{\includegraphics*{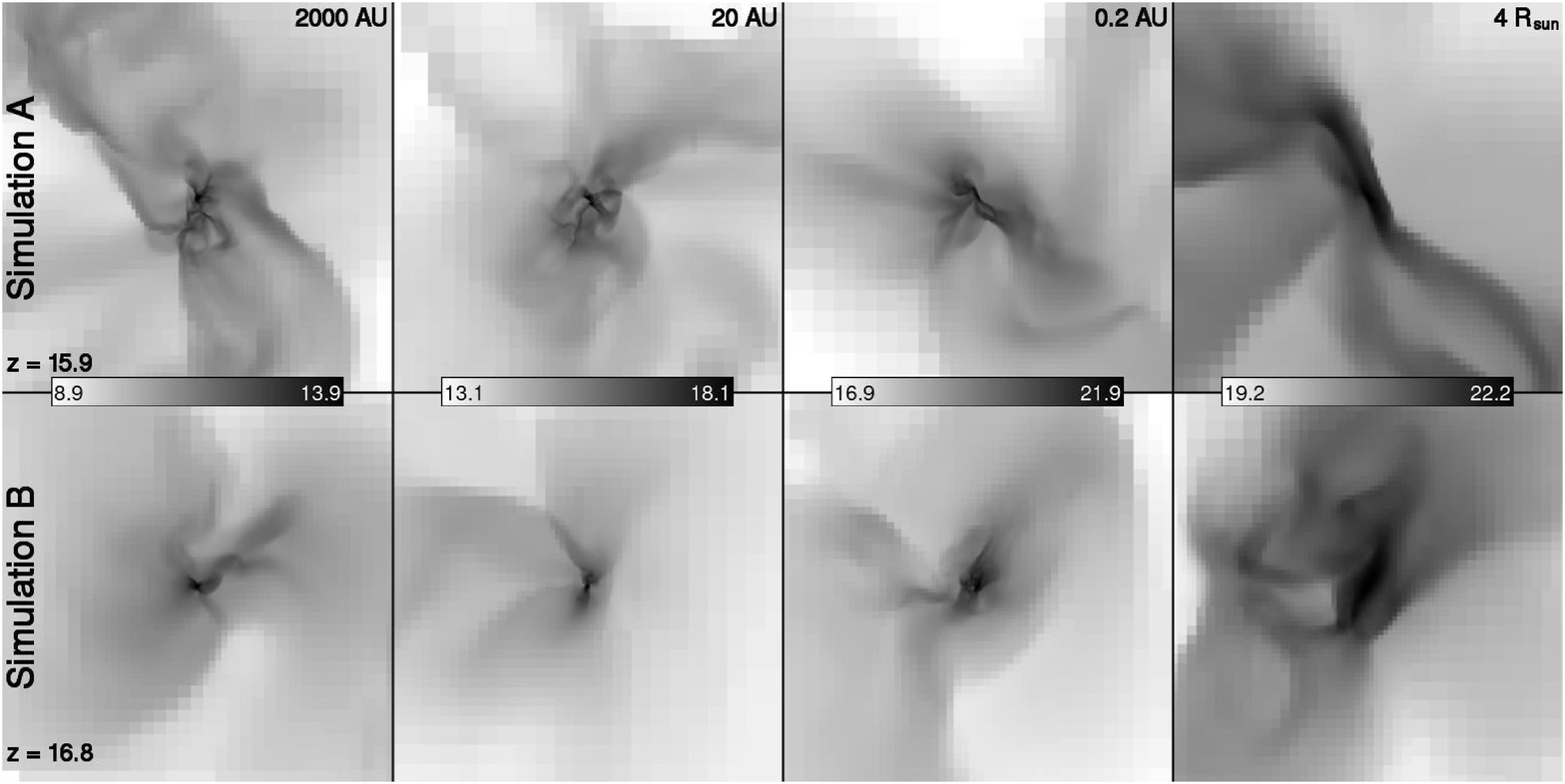}}}
  \caption{An overview of the final state of the collapsing
    protogalactic gas cloud.  Slices of log gas density in cm$^{-3}$
    are shown through the densest point in the halo.  The
    \textit{first} and \textit{three} rows show simulation A, and the
    \textit{second} and \textit{fourth} rows show simulation B.  The
    columns in the top two rows from left to right are slices with a
    field of view of 10 kpc, 1 kpc, 100 pc, and 1 pc.  For the bottom
    two rows, the fields of view are 0.01pc, 20AU, 0.2AU, and 4
    R$_\odot$.  Note that each color scale is logarithmic, spans 5
    orders of magnitude, and is unique for every length scale.}
  \label{fig:slices} 
\end{figure*}

\section{SIMULATION TECHNIQUES}

To investigate protogalactic halo collapses in the early universe, we
utilize an Eulerian structured, adaptive mesh refinement (AMR),
cosmological hydrodynamical code, \enzo\footnote{See
  http://lca.ucsd.edu/software/enzo/} \citep{Bryan97, Bryan99,
  OShea04}.  \enzo~solves the hydrodynamical equations using a second
order accurate piecewise parabolic method \citep{Woodward84, Bryan94},
while a Riemann solver ensures accurate shock capturing with minimal
viscosity.  Additionally \enzo~ uses a particle-mesh N-body method to
calculate the dynamics of the collisionless dark matter particles
\citep{Couchman91}.  Regions of the simulation grid are refined by a
factor of two when one or more of the following conditions are met:
(1) Baryon density is greater than 3 times $\Omega_b \rho_0
N^{l(1+\phi)}$, (2) DM density is greater than 3 times
$\Omega_{\rm{CDM}} \rho_0 N^{l(1+\phi)}$, and (3) the local Jeans
length is less than 16 cell widths.  Here $N = 2$ is the refinement
factor; $l$ is the AMR refinement level; $\phi = -0.3$ causes more
frequent refinement with increasing AMR levels, i.e. super-Lagrangian
behavior; $\rho_0 = 3H_0^2/8\pi G$ is the critical density; and the
Jeans length, $L_J = \sqrt{15kT/4\pi\rho G \mu m_H}$, where $H_0$,
$k$, T, $\rho$, $\mu$, and $m_H$ are the Hubble constant, Boltzmann
constant, temperature, gas density, mean molecular weight in units of
the proton mass, and hydrogen mass, respectively.  The Jeans length
refinement insures that we meet the Truelove criterion, which requires
the Jeans length to be resolved by at least 4 cells on each axis
\citep{Truelove97}. Runs with a refinement criterion of 4, 8, and 16
Jeans lengths have indistinguishable mass weighted radial profiles.

We conduct the simulations within the concordance $\Lambda$CDM model
with WMAP 1 year parameters of $h$ = 0.72, \Ol~= 0.73, \Om~= 0.27,
\Ob~= 0.024$h^{-2}$, and a primordial scale invariant ($n$ = 1) power
spectrum with $\sigma_8$ = 0.9 \citep{Spergel03}.  $h$ is the Hubble
parameter in units of 100 km s$^{-1}$ Mpc$^{-1}$.  \Ol, \Om, and
\Ob~are the fractions of critical energy density of vacuum energy,
total matter, and baryons, respectively.  $\sigma_8$ is the rms of the
density fluctuations inside a sphere of radius 8$h^{-1}$ Mpc.  Using
the WMAP1 parameters versus the significantly different WMAP third
year parameters \citep[WMAP3;][]{Spergel07} have no effect on the
evolution of individual halos that are considered here.  At high
redshifts, statistical differences in structure formation within WMAP3
cosmology when compared to WMAP1 are primarily caused by less
small-scale power prescribed by the lower $\sigma_8$ value (0.9
$\rightarrow$ 0.76) and scalar spectral index $n$ (1 $\rightarrow$
0.96) of primordial density perturbations.  This manifests in (1) a
time delay of $\sim$$40\%$ of the halo formation times for a given
virial mass \citep{Alvarez06}, (2) a corresponding lower halo
abundance for star-forming halos \citep{Gao07, Wang08}, and (3)
stronger clustering of halos \citep{Wang08}.  The initial conditions
of this simulation are well-established by the primordial temperature
fluctuations in the cosmic microwave background (CMB) and big bang
nucleosynthesis (BBN) \citep[][and references therein]{Burles01,
  Hu02}.

We perform two realizations in which we vary the box size and random
phase to study different scenarios and epochs of halo collapse.  In
the first simulation, we setup a cosmological box with 1 comoving Mpc
on a side (simulation A), periodic boundary conditions, and a 128$^3$
top grid.  The other simulation is similar but with a box side of 1.5
comoving Mpc and a different random phase (simulation B).  We provide
a summary of the simulation parameters in Table \ref{tab:params}.
These volumes are adequate to study halos of interest because the
comoving number density of $>$10$^4$ K halos at $z=10$ is $\sim$6
Mpc$^{-3}$ according to an ellipsoidal variant of Press-Schechter
formalism \citep{Sheth02}.  We use the COSMICS package to calculate
the initial conditions%
\footnote{To simplify the discussion, simulation A will always be
  quoted first with the value from simulation B in parentheses.} at
$z$ = 129 (119) \citep{Bertschinger95, Bertschinger01}.  It calculates
the linearized evolution of matter fluctuations.  We first run a dark
matter simulation to $z=10$ and locate the DM halos using the HOP
algorithm \citep{Eisenstein98}.  We identify the first dark matter
halo in the simulation that has \tvir~$>$ 10$^4$ K and generate three
levels of refined, nested initial conditions with a refinement factor
of two that are centered around the Lagrangian volume of the halo of
interest.  The nested grids that contain finer grids have 8 cells
between its boundary and its child grid.  During the simulation, the
initial grids retain its position and are always refined to its
initial resolution or higher.  Their boundary conditions with each
other are treated as any other adaptive grid.  The finest grid has an
equivalent resolution of a 1024$^3$ unigrid and a side length of 250
(300) comoving kpc.  This resolution results in a DM particle mass of
30 (101) $\Ms$ and an initial gas resolution of 6.2 (21) $\Ms$.  These
simulations continue from the endpoints of simulations A6 and B6 of
Paper I.  Table \ref{tab:runs} lists the parameters of the most
massive halo in each realization.  We evolve the system until the
central object has collapsed and reached our resolution limit.  If we
were to follow the simulation to later times and focus on subsequently
collapsing halos, the nature of the gaseous collapses in these halos
should be similar because we do not consider any non-local feedback
processes that affect neighboring halos.  At redshift 15, the mean
separation of halos with \tvir~$>$ 10$^4$ K is 540 and 910 comoving
kpc in WMAP1 and WMAP3 cosmology, respectively, using Sheth-Tormen
formalism \citep{Sheth02}.  Thus we argue that the results presented
here should be applicable to all high-redshift protogalactic
collapses.

There are 1.23 $\times$ 10$^8$ (498$^3$) and 7.40 $\times$ 10$^7$
(420$^3$) unique cells in the final simulation output of simulations A
and B, respectively.  The finest grid then has a refinement level of
41 and a spatial resolution of roughly 0.01 of a solar radius in both
simulations.

\enzo~employs a non-equilibrium chemistry model \citep{Abel97,
  Anninos97}, and we consider six species in a primordial gas (H,
H$^{\rm +}$, He, He$^{\rm +}$, He$^{\rm ++}$, e$^{\rm -}$).  Compton
cooling and heating of free electrons from the CMB and radiative losses
from atomic cooling are computed in the optically thin limit.  At high
densities in the halo cores, the baryonic component dominates the
material.  However, the discrete sampling of the DM potential by
particles can become inadequate, and artificial heating (cooling) of
the baryons (DM) can occur.  To combat this effect, we smooth the DM
particles in cells with a width $<$0.24 ($<$0.36) comoving pc, which
corresponds to a refinement level of 15.

\vspace{1em}
\section{RESULTS} 

In this section, we first describe how the halo collapses when it
starts to cool through \lya~line emission.  Then we discuss the role
of turbulence in the collapse.  Last we describe the rotational
properties and stability of the halo and central object.

%
%

\begin{figure*}
  \includegraphics[width=0.48\textwidth]{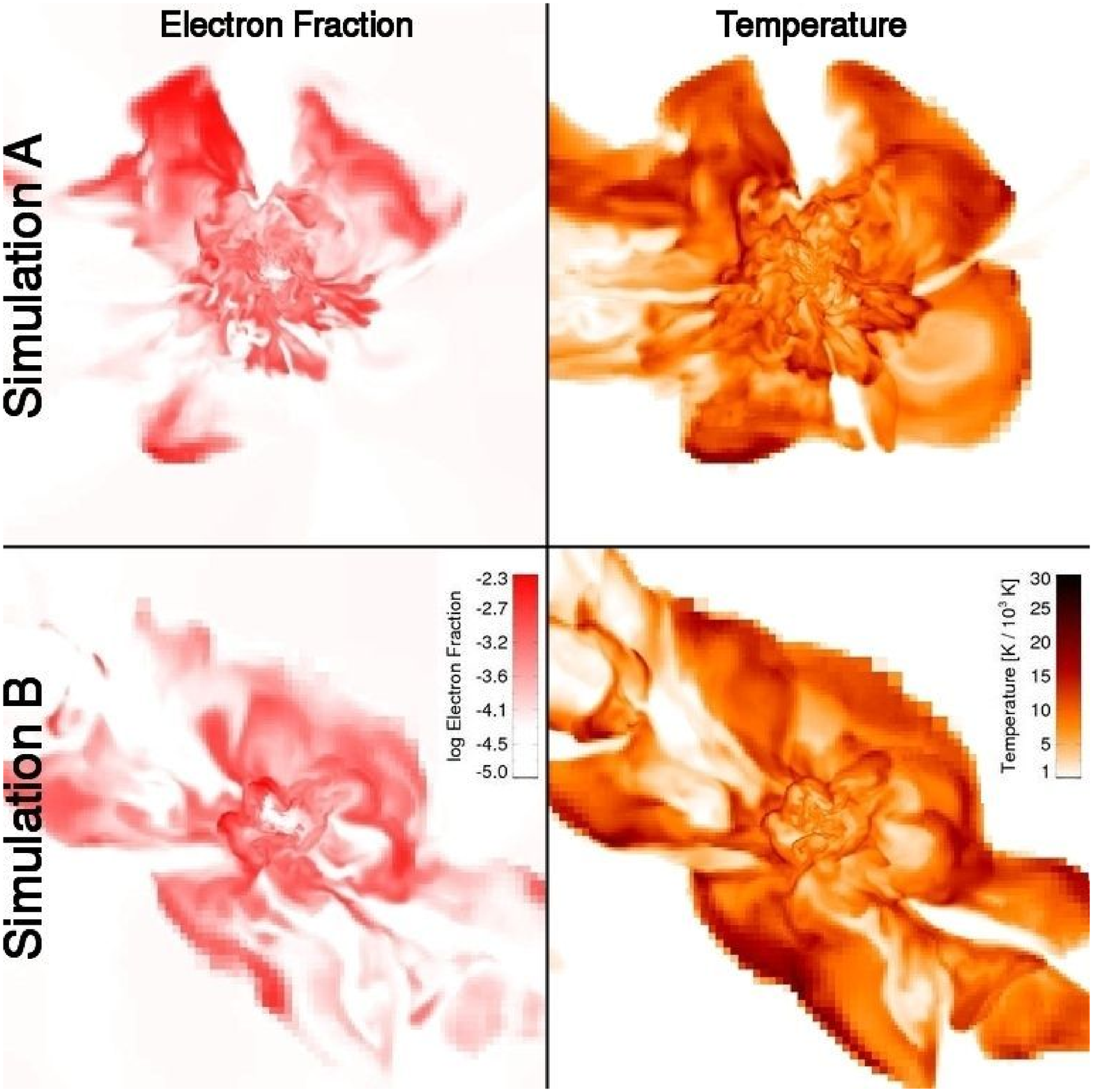}
  \hspace{0.025\textwidth}
  \includegraphics[width=0.48\textwidth]{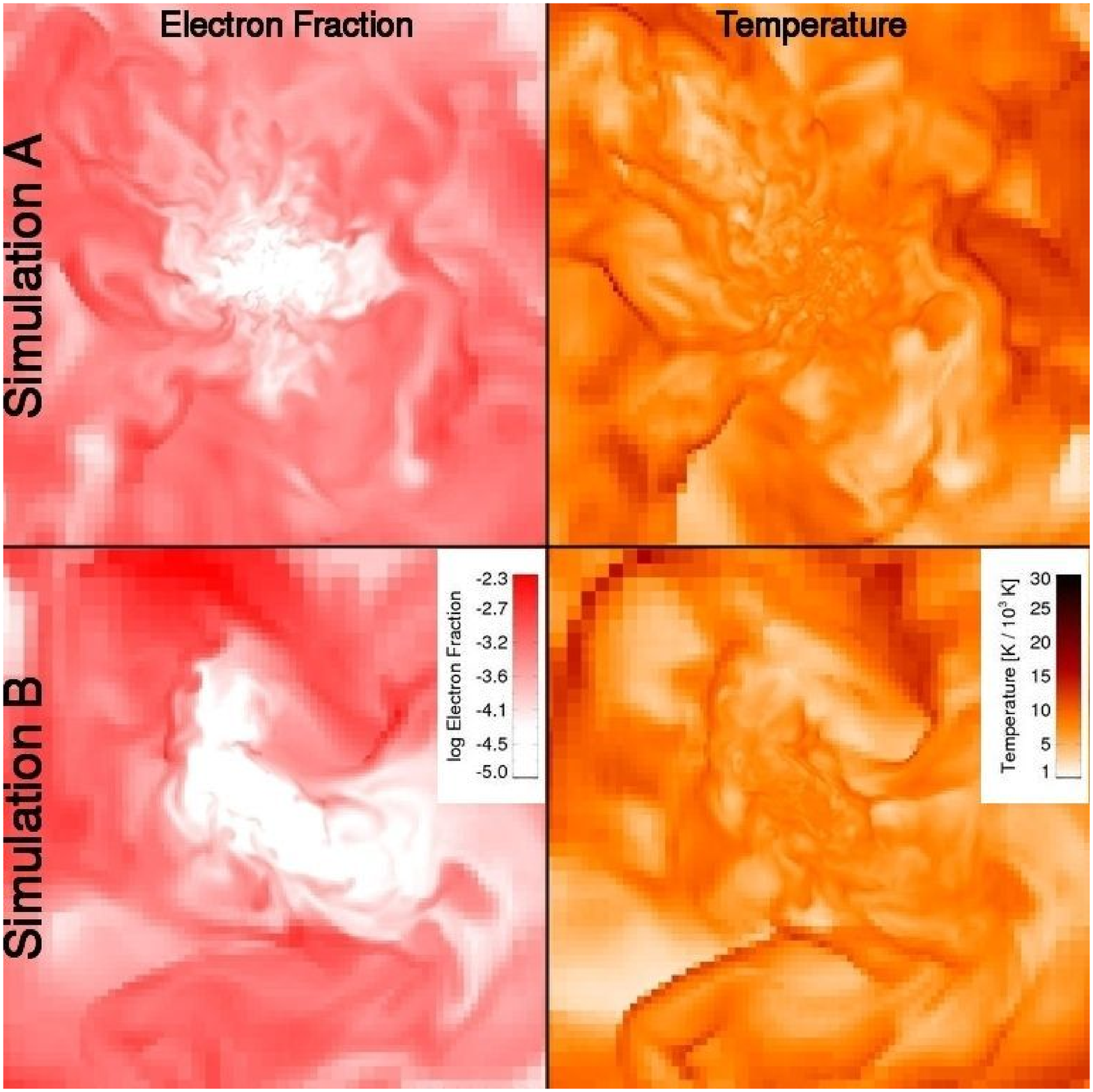}
  \caption{Slices of electron fraction (\textit{left}) and temperature
    (\textit{right}) of simulation A (\textit{top}) and B
    (\textit{bottom}).  The field of view is 1.5 kpc (\textit{left
      panels}) and 200 pc (\textit{right panels}).  The color scale is
    logarithmic for electron fraction and linear for temperature in
    units of 10$^3$ K.  Supersonic turbulent shocks are ubiquitous
    throughout the halos.}
  \label{fig:tempElec1}
\end{figure*}

\subsection{Halo Collapse}
\label{sec:collapse}


Beginning at z = 21.1 in simulation A, the progenitor of the final
halo (\mvir~= 4.96 $\times$ 10$^6 \Ms$) starts to experience two major
mergers, which continues until z = 17.2 when \mvir~= 2.36 $\times$
10$^7 \Ms$.  We define \mvir~as the mass M$_{200}$ in a sphere that
encloses an average DM overdensity of 200.  In simulation B, no recent
major merger occurs before the cooling gas starts to collapse, but it
accumulates mass by accretion and minor mergers.

Mergers disrupt the relaxed state of the progenitor and create
turbulence as these systems collide and combine.  Additional turbulence
arises during virialization, as discussed in Paper I.  More small
scale density fluctuations are thus present in simulation A.  These
fluctuations penetrate farther into the potential well in simulation A
to scales%
\footnote{Note that all masses concerning the collapse are gas mass,
  not total mass.  The central regions of r $<$ 10 pc are baryon
  dominated so that $M_{{\rm enc,\; gas}} \approx M_{{\rm enc,\;
      tot}}$.  All length scales are in proper units unless otherwise
  noted.} of 1 pc, compared to simulation B that contains nearly no
fluctuations between 1 and 50 pc.  This is apparent in the $l$ = 1 pc
panels of Figure \ref{fig:slices} that show the density slices at
eight length scales covering 11 orders of magnitude.  At the 10 kpc
scale, the filamentary large-scale structure is shown, and the
protogalactic halo exists at the intersection of these filaments.  In
the next scale, we show the protogalactic gas cloud.  At the 100 pc
scale, a thick disc is seen in simulation B.  It is nearly edge-on and
oriented northwest to southeast in this view.  In simulation B at 1
pc, a bar forms from a rotational secular instability that transports
angular momentum outwards.  Similar instabilities exist at radii of
0.2 pc, 2700 AU, 17 AU, 0.5 R$_\odot$ in simulation B.  Simulation A
also undergoes a secular bar instability at smaller scales at radii of
150 AU, 1.3 AU, 0.8 R$_\odot$ but shows a more disorganized medium at
larger scales.

The virial temperatures are now $\geq 10^4$ K, and therefore they can
efficiently cool by atomic hydrogen transitions.  The gas fulfills the
critical condition for contraction, \tdyn~$>$ \tcool, and proceeds to
continuously collapse on approximately a dynamical time.  We note that
this collapse and level of fragmentation are strongly influenced by
the magnitude of radiative cooling that the gas can acheive.  Here we
present the case in which the gas cools without any external radiation
backgrounds or radiation trapping, which may alter the nature of the
collapse.

%
%

\begin{figure}[b]
  \resizebox{\columnwidth}{!}{\includegraphics*{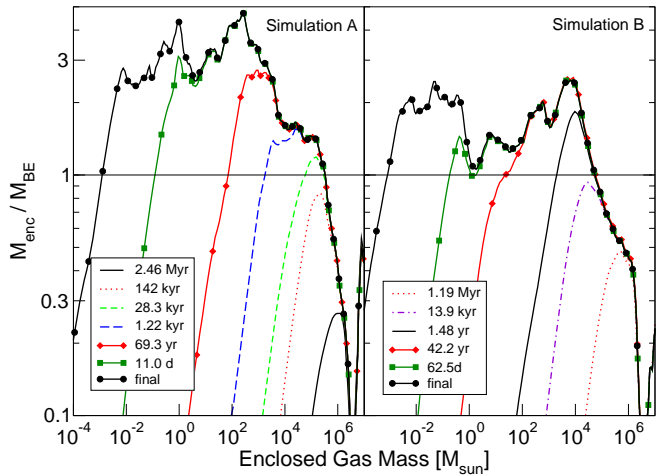}}
  \caption{The ratio of the enclosed gas mass and Bonnor-Ebert mass
    (eq. \ref{eqn:mbe}) for the final output {\em (black with
      circles)} and selected previous times that are listed in the
    legend.  Simulation A (\textit{left}) and B (\textit{right}).  For
    values above the horizontal line at $M_{\rm{enc}} / M_{\rm{BE}} =
    1$, the system is gravitationally unstable.}
  \label{fig:mbe}
\end{figure}

%
%

\begin{figure*}
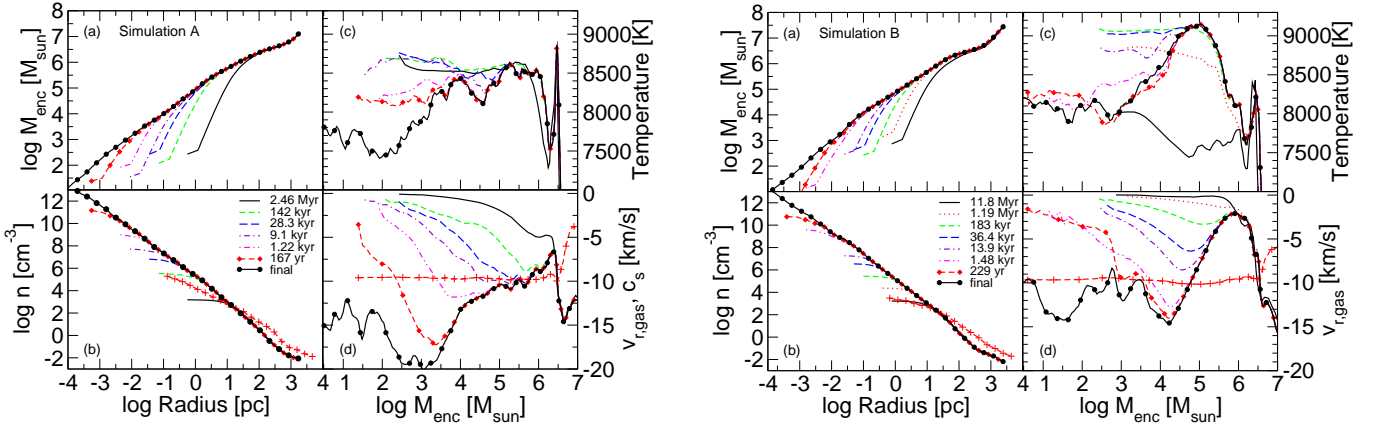

\resizebox{0.48\textwidth}{!}{\rotatebox{0}{\includegraphics*{f4a_color}}}
\hspace{0.025\textwidth}
\resizebox{0.48\textwidth}{!}{\rotatebox{0}{\includegraphics*{f4b_color}}}
\caption{Mass-weighted radial profiles at various times of (a) gas
  mass enclosed, (b) number density, (c) mass-weighted temperature,
  and (d) mass-weighted radial velocity for simulation A (\textit{left
    panels}) and simulation B (\textit{right panels}).  The quantities
  in the left and right panels are plotted with respect to radius and
  gas mass enclosed, respectively.  In (b), the dashed line with
  crosses is the dark matter density in units of $m_H$ cm$^{-3}$.  In
  (d), the dashed line with crosses is the negative of the sound speed
  in the final output.  The times in the legends correspond to time
  before the end of the simulation.}
\label{fig:profilesA} 
\end{figure*}

%
%
%

\begin{figure*}
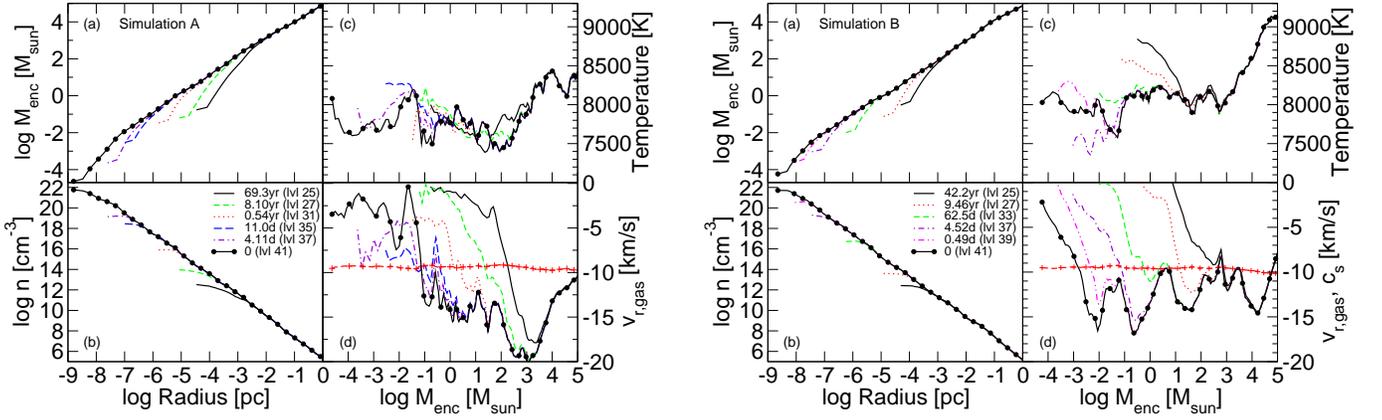

\resizebox{0.48\textwidth}{!}{\rotatebox{0}{\includegraphics*{f5a_color}}}
\hspace{0.025\textwidth}
\resizebox{0.48\textwidth}{!}{\rotatebox{0}{\includegraphics*{f5b_color}}}
\caption{Same as Figure \ref{fig:profilesA} for the inner parsec of
  simulation A (\textit{left panels}) and simulation B (\textit{right
    panels}).  The maximum AMR level is listed next to the times in
  the legend.  In simulation B, the local minima in radial velocities
  at $2 \times 10^4$, 40, 0.3, and 0.01 $\Ms$ occur as angular
  momentum is transported outwards in secular bar-like
  instabilities.}
\label{fig:profilesA3} 
\end{figure*}

Figure \ref{fig:tempElec1} depicts slices of electron fraction and gas
temperature at scales of 200 and 1500 pc.  At the larger scale, the
gas is heated both in virial shocks at $r \sim 600$ pc and internal
turbulent shocks.  Gas within the virial radius varies between
$\sim$2000 K in cold inflows from filaments and up to 30,000 K in
turbulent shocks.  Electron fractions increase to up to 0.5\% because
of collisional ionizations behind the shocks.  The majority of the
ionizations occur in the turbulent shocks inside \rvir~where the
densities are greater and temperatures at the shocks are similar to
values in the virial shock.  In addition, 84\% of the cooling
radiation originates in converging flows ($\nabla \cdot v < 0$).  In
the inner 200 pc, turbulent shocks are widespread as seen in the
temperature variations.  However these are less pronounced than the
one at larger radius.  In the central 50 pc, the gas becomes nearly
isothermal despite the low free electron fraction.

The halo collapses in two stages.  We denote the beginning of the
first stage when \tdyn~$>$ \tcool~for the first time.  The second
stage begins when the central object becomes gravitationally unstable.

1. \textit{Cooling stage}--- As mass infalls toward the center, the
increased cooling rate, which is $\propto nn_e$ until \lya~radiation
becomes trapped within the inner condensation at a density of $\sim 5
\times 10^8 \cubecm$ \citep{Oh02}, catalyzes the collapse as atomic
line transitions convert kinetic energy to radiation.  Here $n$ and
$n_e$ are the number density of baryons and electrons, respectively.
Although we do not treat the radiative effects of \lya, radiation
trapping from recombination lines cannot prevent the collapse
\citep{Rees77}.  This first stage starts 520 (40) kyr before the last
output.  The inner 100 pc have a steady decrease in electron fraction
that indicates atomic hydrogen cooling is now efficient in this
region, which can be seen in the 200 pc slices of Figure
\ref{fig:tempElec1}.  However, only the gas within 1.5 (1.0) pc has
\tdyn~$\gsim$ \tcool~= 383 (100) kyr at this epoch.

2. \textit{Gravitationally unstable stage}--- This starts when the
central region becomes unstable to gravitational collapse.
\citet{Ebert55} and \citet{Bonnor55} investigated the stability of an
isothermal sphere with an external pressure $P_{ext}$ and discovered
that the critical mass (BE mass hereafter) for gravitational collapse
is
\begin{equation}
\label{eqn:mbe}
M_{{\rm BE}} = 1.18 \frac{c_s^4}{G^{3/2}} P_{ext}^{-1/2} \,\Ms .
\end{equation}
If we set $P_{ext}$ to the local pressure, then
\begin{equation}
M_{{\rm BE}} \approx 20 T^{3/2} n^{-1/2} \mu^{-2} \gamma^2 \Ms ,
\end{equation}
where $\gamma$ = 5/3 is the adiabatic index.  For both simulations,
this stage occurs between 10 and 100 kyr before we end the simulation.
We plot the ratio of the enclosed gas mass and BE mass in Figure
\ref{fig:mbe} for several epochs in the collapse.  When the clump
becomes gravitationally unstable, the central 3.3 $\times$ 10$^5$ (5.5
$\times$ 10$^4$) $\Ms$ in the central $r_{\rm{BE}}$ = 5.8 (0.9) pc
exceeds the BE mass, and its \tdyn~= 520 (80) kyr.  Thus our numerical
results agree with these analytic expectations.

We follow the evolution of the accretion and contraction until the
simulation\footnote{We stop the simulation because of ensuing
  round-off errors from a lack of precision.  We use 80-bit precision
  arithmetic for positions and time throughout the calculation.}
reaches a refinement level of 41 (41) that corresponds to a resolution
of 0.01 (0.014) R$_\odot$.  At this point, the central 4.7 $\times$
10$^5$ (1.0 $\times$ 10$^5$) $\Ms$ are gravitationally unstable and
\textit{not} rotationally supported.  The central mass is nearly
devoid of free electrons where the electron fraction, $n_e / n <
10^{-6}$, and the temperature is $\sim8000$ K.  It has a radius of 7.9
(1.5) pc.  The central number density is 5.8 (7.6) $\times$ 10$^{21}$
cm$^{-3}$.  We repeat that this isothermal collapse occurs through
atomic hydrogen cooling only, but in reality, \hh~cooling is important
even in the presence of a ultraviolet background
\citep[e.g.][]{Machacek01, Wise07b, OShea08}.  Thus our results should
only be considered as a scenario for excellent numerical experiments
of turbulent collapses (see \S\ref{sec:applicability} for more
discussion).

Next we show the radial profiles of the final and preceding outputs in
Figures \ref{fig:profilesA} and \ref{fig:profilesA3}, where we plot
(a) enclosed gas mass, (b) number density, (c) mass-weighted
temperature, and (d) mass-weighted radial velocity.  Figure
\ref{fig:profilesA} focuses on length scales greater than 20 AU to $r
> \rvir$.  The halo collapses in a self-similar manner with $\rho(r)
\propto r^{-12/5}$.  We also overplot the DM density in units of
m$_{\rm{H}} \cubecm$ in the $b$ panels.  The DM density in simulation
A does not flatten as much as simulation B with $\rho_{\rm{DM}}
\propto r^{-4/3}$ and $r^{-2/3}$, respectively, yet higher DM
resolution simulations will be needed to address the significance of
this difference in central slopes.  In the $c$ panels, ones sees that
the entire system is isothermal within 10\% of 8000 K. In the $d$
panels, the sound speed $c_s$ in the final epoch is plotted, and there
is a shock where $v_r > c_s$ at a mass scale when $M_{{\rm enc}}$
first exceeded $M_{{\rm BE}}$.  Here $v_r$ is the radial velocity, and
$c_s$ is the local sound speed.

Figure \ref{fig:profilesA3} shows the data within 1 pc at times 100
years before the end of the simulation.  The self-similar, isothermal
collapse continues to stellar scales.  However, the structure in the
radial velocity in simulation B exhibits a strikingly behavior with
four repeated minima at mass scales $2 \times 10^4$, $10^3$, 6, and
$10^{-3} \Ms$.  We attribute this to rotational bar-like instabilities
that we discuss later in the paper (\S\ref{sec:rotInstab}).

\begin{figure}
\resizebox{\columnwidth}{!}{\rotatebox{0}{\includegraphics*{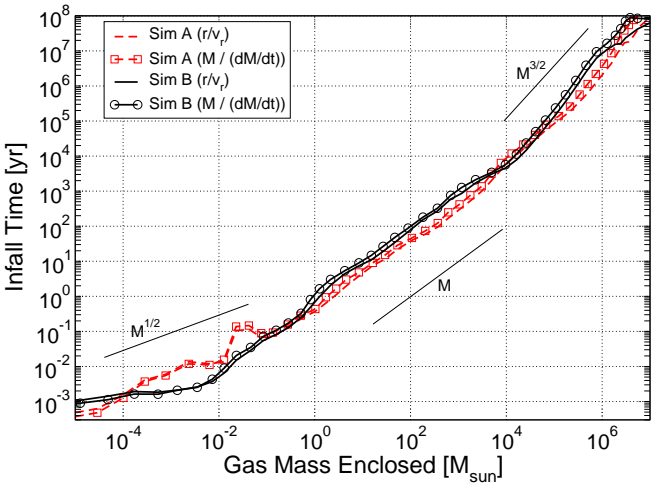}}}
\caption{Radial profiles of gas infall times at the final output.  To
  approximate a collapse timescale, the quantities $r/v_r$ {\em
    (solid)} and $M/(dM/dt)$ {\em (dashed)} are calculated and plotted
  here.}
\label{fig:infall} 
\end{figure}

If we consider $v_r$ constant from the last output, we can determine
the infall times, which are shown in Figure \ref{fig:infall}.  The
infall time, $t_{in} = r/v_r$, of the shocked BE mass is 350 (50) kyr.
The infall times approximately follow a broken power law, $t_{in}
\propto M_{\rm{enc}}^\beta$.  Within $M_{\rm{enc}} \sim 0.1 \Ms$,
$\beta \approx 1/2$.  In the range $0.1 \lsim M_{\rm{enc}}/\Ms \lsim 3
\times 10^4$, $\beta \approx 1$; above this mass interval, the slope
of the mass infall times increase to $\beta \approx 3/2$.  The
increased radial velocities when the central object becomes
gravitationally unstable causes the steepening of the slope at
$\sim$$3 \times 10^4 \Ms$.

%
%

\vspace{1em}
\subsection{Global Disc}

In simulation B, a thick disc with a radius of 50 pc and disc scale
height of $\sim$10 pc forms that is pressure supported and only
partially rotationally supported.  The circular velocities within this
disc achieve only a third of Keplerian velocities.  The lack of full
rotational support and large scale height suggests that a central
collapse occurs before any fragmentation in this large-scale disc is
possible.  In contrast, we see a disorganized, turbulent medium and no
large scale disc formation in simulation A.

\subsection{Turbulence}
\label{sec:turbulence}

\citet{Kolmogorov41} described a theory of the basic behavior of
incompressible turbulence that is driven on a large scale and forms
eddies at that scale.  These eddies then interact to form smaller
eddies and transfer some of their energy to smaller scales.  This
cascade continues until energy is dissipated through viscosity.  In
supersonic turbulence, most of the turbulent energy is dissipated
through shock waves, which minimizes the local nature of cascades
found in incompressible turbulence.

In Paper I, we found that turbulence is stirred during virialization.
When radiative cooling is efficient, the gas cannot virialize by
gaining thermal energy and must increase its kinetic energy in order
to reach equilibrium, which it achieves by radial infall and turbulent
motions.

In addition to virial turbulence generation, mergers stir turbulence.
Here the largest driving scale will be approximately the scale of the
merging objects, and the turbulent cascade starts from that length
scale.  Additional driving may come from Kelvin-Helmholtz
instabilities of a multi-phase gas as the mergers occur
\citep{Takizawa05}.  \citeauthor{Takizawa05} considered mergers of
galaxy clusters, however his work may still apply to the formation of
protogalactic halos since similar temperature contrasts exist in this
regime of mergers.  As the lighter halo falls into the massive halo, a
bow shock and small-scale eddies from the Kelvin-Helmholtz instability
form between the two interacting objects.  At later times, a dense,
cool core remains in the substructure of the lesser halo.  The
instabilities grow and destroy the baryonic substructure, and the gas
mixes with the existing gas of the massive halo and becomes turbulent.

%
%
%

\begin{figure}
  \resizebox{\columnwidth}{!}{\rotatebox{0}{\includegraphics*{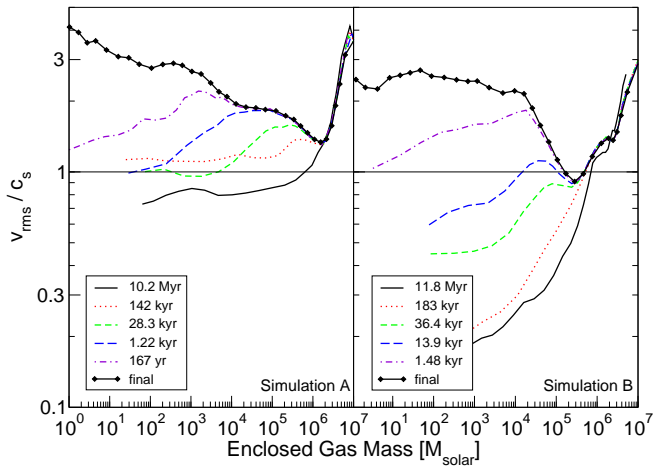}}}
  \caption{The turbulent Mach number, $v_{rms} / c_s$, for the final
    output {\em (black with diamonds)} and selected previous times that
    are listed in the legend.  Simulation A (\textit{left}) and B
    (\textit{right}).}
  \label{fig:mach}
\end{figure}

To quantify aspects of this turbulence, we inspect the turbulent Mach
number,
\begin{equation}
\mathcal{M} = \frac{v_{rms}}{c_s}; \quad 
c_s^2 = \frac{dP}{d\rho} = \frac{\gamma k T}{\mu m_H}.
\end{equation}
Here $P$ is pressure, $v_{rms}$ is the 3D velocity dispersion, and
$\gamma$ is the adiabatic index that we set to 5/3.  We evaluate
$v_{rms}$ with respect to the mean velocity of each spherical shell.
Radial profiles of $\mathcal{M}$ are shown in Figure \ref{fig:mach}.
Before the core becomes gravitationally unstable, the turbulence is
subsonic within the virial shock.  After the core becomes
gravitationally unstable, the turbulent Mach number rises to 2--4.
The collapse produces turbulence on a timescale that is faster than it
can be dissipated.



The turbulence that exists before the initial collapse may impact the
nature of the central object.  In simulation A, the core initially has
$\mathcal{M} \approx 1$, and this results in a central object with 4.7
$\times$ 10$^5 \Ms$ and a radius of 7.9 pc.  The core in simulation B
has $\mathcal{M} \approx 0.2$, and the central object is about five
times less massive and smaller, which corresponds to a free-fall time
approximately five times shorter as well.

\subsection{Spin Parameter Evolution}

\begin{figure}
  \resizebox{\columnwidth}{!}{\rotatebox{0}{\includegraphics*{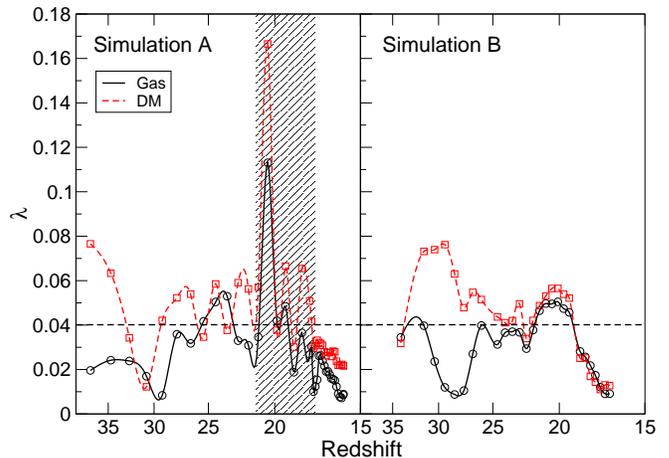}}}
  \caption{Spin parameter, $\lambda \equiv \vert L \vert \sqrt{\vert E
      \vert} / GM^{5/2}$, evolution of the main halo in the
    simulation.  {\em (left)} simulation A.  {\em (right)} simulation
    B.  The dashed and solid lines are the interpolated values for the
    DM and baryonic spin parameter.  The squares and circles
    correspond to the actual measurements from the DM and gas data,
    respectively.  The horizontal dashed line at $\lambda$ = 0.04
    marks the mean cosmological spin parameter.  In simulation A, two
    major mergers causes the large increase beginning at z $\approx$
    21 in the hashed region.  The oscillations occur as the merging
    halos orbit each other until they virialize.}
  \label{fig:spin}
  \vspace{-0.5em}
\end{figure}

During the hierarchical buildup of structure, tidal forces from
neighboring structures impart angular momentum to a halo, particularly
when its radius is maximal at the turn-around time \citep{Hoyle49,
  Peebles69}.  However in recent years, several groups have recognized
that the mergers may impart a considerable fraction of angular
momentum to the system \citep{Steinmetz95, Gardner01, Vitvitska02,
  Maller02}.  Over many realizations of mergers, the net angular
momentum change would be zero.  In reality, an angular momentum
residual remains after the last major merger occurs because there are
too few events to cancel the randomization of halo spin.  Although
each halo has unique rotational properties, it is useful to define a
dimensionless spin parameter
\begin{equation}
\lambda \equiv \frac{\vert L \vert \sqrt{\vert E \vert}} {G M^{5/2}},
\end{equation}
where G is the gravitational constant and L, E, and M are the angular
momentum, energy, and mass of the object, that measures the rigid body
rotation of the halo \citep{Peebles71}.  In Figure \ref{fig:spin}, we
display the time evolution of $\lambda$ of the DM and baryons in our
simulations and mark the occurrence of the major merger in simulation
A. \citet{Eisenstein95b} \citep[preceded by][]{Barnes87} calculated
that the mean spin parameter, $\langle\lambda\rangle \approx 0.04$, is
weakly dependent on object mass and cosmological model, and this value
is also marked in Figure \ref{fig:spin}.  Also $\lambda$ weakly
depends on its merger history, where $\langle\lambda\rangle$ increases
during mergers and slowly dissipates afterwards.  Most of the angular
momentum is acquired from steady minor mergers and accretion because
major mergers only happen rarely (usually only once per logarithmic
mass interval).  In 96\% of mergers, the majority of the internal spin
originates from the orbital energy of the infalling halo
\citep{Hetznecker06}.

%
%

At $z \approx 22$ in simulation A, the spin parameter $\lambda = 0.06$
before the last major merger.  Then the spin parameter increases by a
factor of 3 during its major merger because of the system being far
from dynamical equilibrium.  The system becomes virialized after
approximately a dynamical time, and the spin parameter stabilizes at
$\lambda \approx 0.03$ and proceeds to decrease with time until
$\lambda = 0.022$ at the time of collapse.  The above evolution of
$\lambda$ agrees with the findings of Hetznecker \& Burkert.
Simulation B describes a halo that does not undergo a recent major
merger, and its final $\lambda$ = 0.013.

Both halos have less angular momentum than $\langle \lambda \rangle$
when the cooling gas collapses.  The probability distribution of
$\lambda$ can be described with the log-normal function
\begin{equation}
  \label{eqn:lambda_prob}
  p(\lambda)d\lambda = \frac{1}{\sigma_\lambda \sqrt{2\pi}} \exp
  \left[ -\frac{\ln^2 (\lambda/\lambda_0)}{2\sigma_\lambda} \right]
  \frac{d\lambda}{\lambda},
\end{equation}
where $\lambda_0 = 0.042 \; \pm \; 0.006$ and $\sigma_\lambda = 0.5 \;
\pm \; 0.04$ \citep[e.g.][]{Bullock01}.  From the cumulative
probability function resulting from equation (\ref{eqn:lambda_prob}),
89\% (99\%) of the cosmological sample of halos have larger spin
parameters than the halos described here.  \citet{Eisenstein95a}
demonstrated that halos with low spin parameters are candidates for BH
formation and quasar seeds.  However they argue that the angular
momentum needs to be at least an order of magnitude lower than the
mean.  Next we present further evidence that reveals a gaseous
collapse is possible with not too atypical spin parameters.

%
%

\begin{figure}
  \resizebox{\columnwidth}{!}{\rotatebox{0}{\includegraphics*{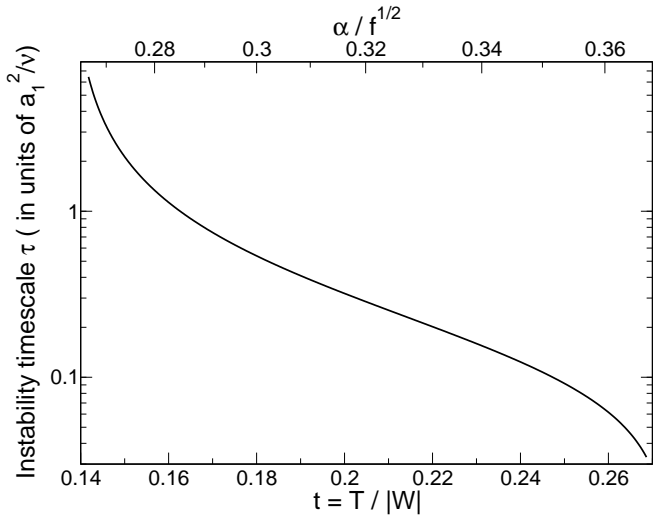}}}
  \caption{Secular instability $e$-folding timescale in units of
    $a_1^2 / \nu$ as a function of t = $T / \vert W \vert$ and $\alpha
    = (tf/2)^{1/2}$ (eq. \ref{eqn:alpha}).  At t $<$ 0.1375, the
    system is stable to all perturbations.  Above $t$ = 0.27, the
    system is dynamically unstable, and this timescale is not
    applicable.}
  \label{fig:secular_tau} 
  \vspace{0.5em}
\end{figure}

\subsection{Instability of Maclaurin Spheroids}
\label{sec:analytics}

The dynamics of rotating systems is a classic topic in astrophysics
(see EFE \S\S1--6).  These self-gravitating systems are susceptible to
two types of instabilities.  Secular instability occurs when small
dissipative forces, e.g. viscosity, amplify perturbations to become
unstable in an otherwise stable inviscid configuration.  Dynamical
(also referred to as ordinary) instability results when some
oscillatory mode exponentially grows with time, regardless of any
dissipative forces.  Here we concentrate on Maclaurin spheroids
relevant for a uniform body rotating with a fixed angular velocity.
Maclaurin spheroids are a special case of Jacobi ellipsoids that are
axisymmetric.  The onset of the $m = 2$ bar-like instability in
gaseous Maclaurin spheroids happens for a given eccentricity,
\begin{equation}
  \label{eqn:eccentricity}
  e = \left( 1 - \frac{a_3^2}{a_1^2} \right)^{1/2} \geq \left\{ 
      \begin{array}{r@{\quad}l}
        0.8127 & \mathrm{(secular)} \\
        0.9527 & \mathrm{(dynamical)}
      \end{array}
  \right.,
\end{equation}
where $a_3$ and $a_1$ are the principle axes with $a_3 \leq a_1$ (EFE
\S33).  Eccentricity is related to the ratio, $t = T / \vert W \vert$,
of rotational kinetic energy to gravitational potential by
\begin{equation}
  \label{eqn:t_vs_e}
  t = \frac{1}{2} [ (3e^{-2} - 2) - 3(e^{-2} - 1)^{1/2} (\sin^{-1}
  e)^{-1} ],
\end{equation}
and the secular and dynamical instabilities happen at $t = (0.1375,
0.27)$, respectively \citep[e.g.][]{Ostriker73b}.

When $t$ is larger than 0.1375 but smaller than 0.27, both the
Maclaurin spheroid and Jacobi ellipsoid are perfectly stable against
small perturbations in the inviscid case.  For a given $e$, the Jacobi
configuration has a lower total energy than its Maclaurin counterpart
and is therefore a preferred state.  Here any dissipative force
induces a secular bar-like instability.  The system slowly and
monotonically deforms through a series of Riemann S-type ellipsoids
until its final state of a Jacobi ellipsoid with an equal angular
momentum \citep{Press73} and lower angular velocity (EFE \S32) as
specific angular momentum is transported outward.  The instability
grows on an $e$-folding timescale
\begin{equation}
\label{eqn:secular_tau}
\tau = \phi a_1^2 / \nu,
\end{equation}
where $\phi$ is a constant of proportionality that asymptotes at $t$ =
0.1375, decays to zero at $t$ = 0.27, and is plotted in Figure
\ref{fig:secular_tau} (EFE \S37).  Here $\nu$ is the kinematic
viscosity.

\citet{Christodoulou95a, Christodoulou95b} generalized the
formulations for bar-like instabilities to account for self-gravity.
In addition, they consider different geometries, differential
rotation, and non-uniform density distributions.  They devised a new
stability criterion
\begin{equation}
  \label{eqn:alpha}
  \alpha \equiv \frac{T/\vert W \vert}{\Omega / \Omega_J} =
  \sqrt{\frac{f}{2} \frac{T}{\vert W \vert}}
\end{equation}
where $\Omega$ is the rotation frequency, 
\begin{equation}
\Omega^2_J = 2 \pi G \rho \left[ \frac{(1-e^2)^{1/2}}{e^3} \sin^{-1} e
  - \frac{1-e^2}{e^2} \right]
\end{equation}
is the Jeans frequency in the radial direction for a Maclaurin
spheroid, and
\begin{equation}
  \label{eqn:rotF}
  f = \frac{1}{e^2} \left[ 1 - \frac{e}{\sin^{-1} e} \sqrt{1 - e^2}
  \right]
\end{equation}
accounts for differing geometries\footnote{See
  \citet{Christodoulou95b} for more generalized geometries.} with $f =
2/3$ for a sphere and $f = 1$ for a disc.  Secular and dynamical
instabilities for Maclaurin spheroids occur above $\alpha$ = (0.228,
0.341), respectively, for $f = 1$.

From N-body simulations of disc galaxies, \citet{Ostriker73b} found
that a massive dark halo with comparable mass to the disc could
suppress secular instabilities.  In the case of a gaseous collapse to
a SMBH however, the baryonic component dominates over the dark matter
component in the central 10 pc.  Secular instabilities cannot be
prevented through this process, which we demonstrate next.

\subsection{Rotational Instabilities}
\label{sec:rotInstab}

\begin{figure}
\resizebox{\columnwidth}{!}{\rotatebox{0}{\includegraphics*{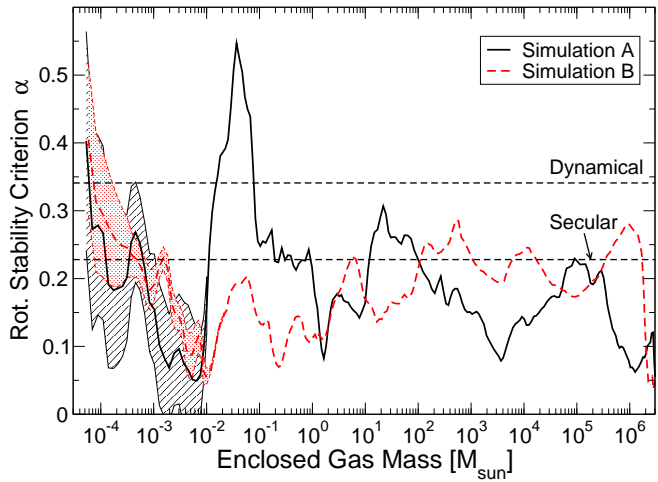}}}
\caption{Rotational instability parameter $\alpha = \sqrt{fT/2\vert W
    \vert}$ for the thick disc with $r \simeq 50$ pc in simulations A
  (\textit{black solid line}) and B (\textit{red dashed line}).  The
  shaded areas show the standard deviation when varying the center on
  the 100 densest points.  For $\alpha > 0.22$ denoted by the
  horizontal line, a secular instability occurs in the disc and leads
  to bar formation.  In simulation A, instabilities occur at mass
  scales of 100, 0.1, and $10^{-4} \Ms$.  In simulation B, the same
  happens at $2 \times 10^6$, $2 \times 10^4$, $10^3$, 6, and $10^{-3}
  \Ms$.  We also mark $\alpha = 0.341$ where a rotating system becomes
  dynamically unstable.  Only simulation A at 0.1 $\Ms$ experiences a
  dynamical instability.}
\label{fig:diskB}
\end{figure}

In the $l$ = 1 pc panel of simulation B in Figure \ref{fig:slices}, it
is apparent a bar-like instability exists in the gravitationally
unstable central object.  Figure \ref{fig:diskB} shows the instability
criterion $\alpha$ (eq. \ref{eqn:alpha}) against enclosed gas mass.
Here we transform the velocities to align the $z$-axis with the
baryonic angular momentum vector of the entire halo.  We use the
tangential velocities to calculate the rotational kinetic energy $T$.
The shape parameter $f$ = 2/3 (0.89) for simulation A (B).

%
%

As discussed before, Maclaurin spheroids are subject to secular $m =
2$ bar-like instabilities when $\alpha > 0.228$.  In simulation A, the
central object becomes unstable at three approximate mass scales, $6.7
\times 10^{-4}$, 1.0, and 110 $\Ms$ that correspond to radii of 0.75
R$_\odot$, 1.3 AU, and 150 AU, respectively.  The enclosed mass ratios
of the recurring instabilities, i.e. $M_i/M_{i+1}$, are 1500:1,
110:1, and 1400:1, starting at the smallest mass scale.  The
instability at 0.075 $\Ms$ ($r$ = 0.13 AU) is dynamically unstable
with $\alpha$ peaking at 0.55.  In simulation B, instabilities occur
at $5.3 \times 10^{-4}$, 7.0, $1.2 \times 10^3$, and $2.0 \times 10^4
\Ms$ at radii of 0.49 R$_\odot$, 17 AU, 2700 AU, and 0.18 pc.  The
enclosed mass ratios of these instabilities, are 13,000:1, 170:1,
17:1, and 85:1.

It is interesting to note that the innermost instability in both
simulations becomes dynamical ($\alpha > 0.341$), and $\alpha$
continues to increase rapidly toward the center.  However these
features should be taken with caution since it occurs near our
resolution limit, where the particular location used as the center
will influence the rotational energy one would calculate.  To evaulate
the sensitivity in choosing a center, we performed the same analysis
but varying the center over the 100 most densest cells in the
simulation.  We plot the standard deviation of $\alpha$ as the shaded
area in Figure \ref{fig:diskB}.  Inside an enclosed mass of $3 \times
10^{-4} \Ms$, it is $\sim$0.05 but diminishes to less than 0.01 outside
$0.1 \Ms$.

The $e$-folding time of secular instabilities $\tau$ is proportional
to $a_1^2$ (see eq. \ref{eqn:secular_tau}).  Hence small-scale
instabilities collapse on a faster timescale than its parent,
large-scale bar instability.  Turbulent viscosity is the main
dissipative force that drives the instability.  $\tau$ is inversely
proportional to the viscosity.  This further shortens $\tau$ because
supersonic turbulence is maintained to the smallest scales.


%
%

\begin{figure*}
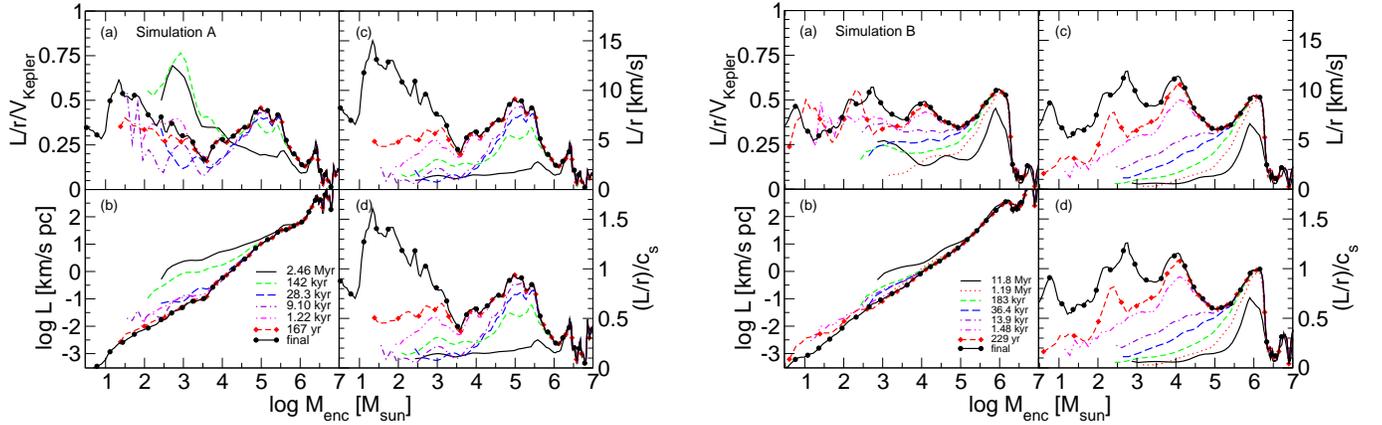

  \resizebox{0.48\textwidth}{!}{\rotatebox{0}{\includegraphics*{f11a_color}}}
  \hspace{0.025\textwidth}
  \resizebox{0.48\textwidth}{!}{\rotatebox{0}{\includegraphics*{f11b_color}}}
  \caption{Mass-weighted radial profiles of various rotational
    quantities in simulation A (\textit{left panels}) and simulation B
    (\textit{right panels}).  In panel A, we show the rotational
    velocity compared to the Kepler velocity = $\sqrt{GM/r}$.  In
    panel B, we display the typical rotational velocity.  In panels C
    and D, the specific angular momentum (in units of km/s pc) and the
    ratio of the rotational velocity and sound speed is shown,
    respectively.  The line styles correspond to the same times in
    Figure \ref{fig:profilesA}.}
  \label{fig:profilesB} 
\end{figure*}

%
%

\begin{figure*}
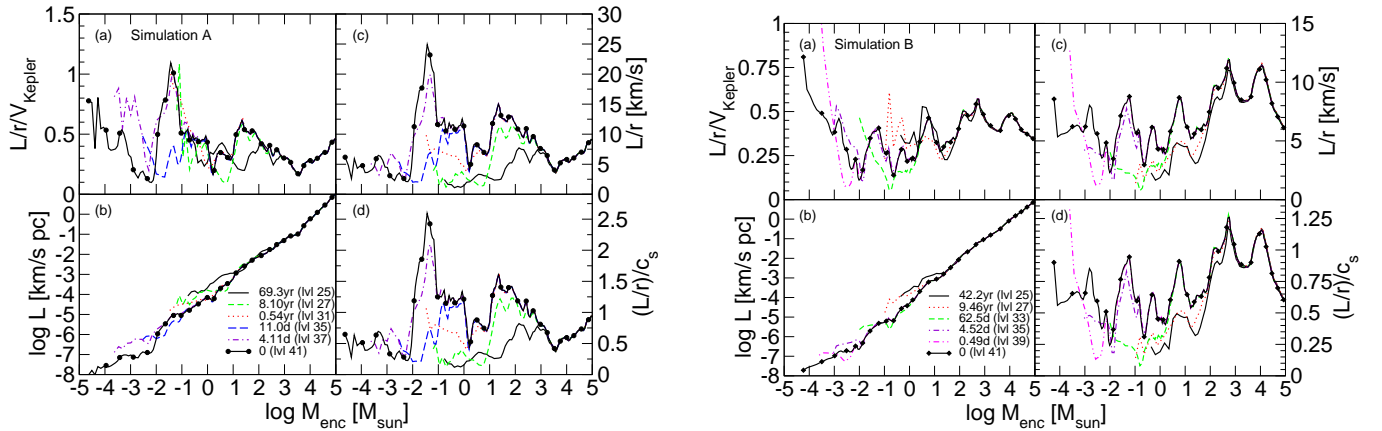

  \resizebox{0.48\textwidth}{!}{\rotatebox{0}{\includegraphics*{f12a_color}}}
  \hspace{0.025\textwidth}
  \resizebox{0.48\textwidth}{!}{\rotatebox{0}{\includegraphics*{f12b_color}}}
  \caption{The same as Figure \ref{fig:profilesB} but with the inner
    parsec of simulation A and B and the output times as listed in
    Figure \ref{fig:profilesA3}.}
  \label{fig:profilesB3} 
\end{figure*}

\subsection{Rotational Properties}

During the collapse of the gas in our simulations, rotational support
never impedes the collapse.  In Figures \ref{fig:profilesB} and
\ref{fig:profilesB3}, we show (a) coherent rotational velocity divided
by Keplerian velocity $v_{kep} = \sqrt{GM/r}$, (b) rotational
velocity, (c) specific angular momentum, and (d) rotational velocity
divided by the sound speed.  We compute the rotational velocities
around the center of mass of a sphere with radius of 100 cell widths
of the finest AMR level, centered on the densest point.  We note that
the rotational velocity $L/r$ plotted here is different than organized
rotation, i.e. a disc.  The radial profiles only sample gas in
spherical shells, whose angular momentum vectors are not necessarily
parallel.

1. \textit{Simulation A}--- At $r > 1$ AU (M$_{\rm{enc}} = 1 \Ms$),
the typical rotational speed is two or three times lower than the
Keplerian velocity, which is required for rotational support.  At $r$
= 0.1 AU (M$_{\rm{enc}}$ = 0.07 $\Ms$), the infall becomes marginally
rotationally supported, i.e. $L/r \sim v_{kep}$.  The radial
velocities react by slowing from 15\kms~to below 5\kms.  However this
rotational support does not continue to the center.  Rotational speeds
are only $\sim$0.5$v_{kep}$ within 0.1 AU (M$_{\rm{enc}} = 1 \Ms$).

2. \textit{Simulation B}--- This collapse exhibits four minima in
radial velocity that are caused by rotational bar-like instabilities.
After such an instability occurs, the radial velocities increase
because of angular momentum being transported outwards.  As the
rotational velocities decrease, this instigates another secular
instability, which repeats causing a cascade of the instability.  The
increased infall velocity and associated decrease in rotational
velocities (i.e. the dips in Figures \ref{fig:profilesA3}d and
\ref{fig:profilesB3}d) depict this behavior.  At the final output, the
infalling material exhibits no rotational support at all radii similar
to simulation A at $r > 1$ AU.

We interpret the inner points where L/r/$v_{\rm{kep}}$ fluctuations
greatly increases above unity with caution because of the nature of
choosing a center in a turbulent medium, i.e. when turbulent
velocities dominate over rotational ones.  If the central sphere is
smaller than a radius where the turbulent velocities average to zero,
we introduce errors into the angular momentum profiles by sampling the
turbulent gas incompletely.  In the $b$-panels of Figure
\ref{fig:profilesB}, one sees that specific angular momentum inside
M$_{\rm{enc}} < 10^6 \Ms$ decreases over time and is transported
outwards in the collapse.




With a not too atypical spin parameter, the thick disc with $r\sim50$
pc is not rotationally supported.  In simulation A, a global disc does
not exist at all.  We attribute this behavior to the nature of angular
momentum transport in a turbulent medium.  Even with a higher spin
parameter, we do not expect a disc to fragment before the central
collapse of gas with low specific angular momentum and short dynamical
times.  This low specific angular momentum material collapses to small
radii without fragmentation so that a central dense object forms with
a mass of $\sim10^5 \Ms$ or 2\% of the halo gas mass.  After the
initial collapse, the thick disc may become rotationally supported as
more high angular momentum gas infalls.

\section{DISCUSSION}
\label{sec:discuss}

In our cosmological simulations, we find that a $\sim$10$^5 \Ms$ dense
object forms in the center of a metal-free protogalactic halo that
cools by atomic hydrogen cooling.  Although we have neglected some
important processes, such as \hh~chemistry, star and BH formation and
feedback, our results show that angular momentum transport at both
small and large scales in the form of preferential segregation and
rotational instabilities, respectively, lead to the formation of a
dense, massive object with r $<$ 5 pc.  This initial central collapse
should precede any fragmentation of a global disc.

\subsection{Angular Momentum Transport}

Collapsing turbulent clouds, whether cosmological or galactic in
nature, are ubiquitous in the universe.  In this paper, we focus on
the details of the turbulent collapse of a proto-galactic halo.
Angular momentum transport plays a key role in such events, e.g.,
determining the characteristics of the central object(s).  However
there exists the ``angular momentum problem'', where many orders of
magnitude of angular momentum must be shed \citep[see \S6
in][]{Larson03} from the initial molecular cloud to form a central
star, star cluster, or BH.  In our simulations, there is a clear
scenario in which the inside-out collapse \citep{Shu87} proceeds even
if the initial turbulent cloud was rotating.  We see three major
elements affecting angular momentum transport during the collapse.

\medskip

1. \textit{Angular momentum distribution}--- In cosmological halos,
there is a universal distribution of angular momentum
\begin{equation}
  M(<j) = M_{\rm{vir}} \frac{\mu j}{j_0 + j}, \quad \mu > 1,
\end{equation}
that measures the mass with a specific angular momentum less than $j$
\citep{Bullock01}.  This function is fitted with two parameters, $\mu$
and $j_0$, where $\mu$ controls the flattening of the power law at
high angular momenta, and $j_0$ determines at which $j$ this
transition occurs.  \citeauthor{Bullock01} also find that more mass
resides in the tails of the distribution, especially at small $j$,
when compared to a system in solid body rotation.  Thus all halos have
some intrinsic amount of gas with small $j$.  If this distribution is
maintained during the collapse \citep[e.g.][]{Mestel63}, such gas can
collapse to some small radius, $r_{\rm{min}} > j/v_{\rm{kep}}$,
without becoming rotationally supported, which leads to the next
element of discussion -- angular momentum segregation.

2. \textit{Segregation in a turbulent medium}--- In Paper I, we
determined that most of the gas becomes supersonically turbulent as a
result of virialization.  Therefore let us theorize how angular
momentum transport happens during the transition from being pressure
supported to rapidly cooling and collapsing.  First consider a
turbulent uniform-density gas cloud, where parcels of gas at a
specific radius can have many different values of $j$.  This differs
from the organized rotation of a disc.  If we start with such an
initial configuration, how does angular momentum transport occur
during the collapse?  Gas with small (high) $j$ will preferentially
migrate to small (large) radii, following turbulent flow lines.  In an
axi-symmetric system, the Rayleigh criterion \citep{Rayleigh20,
  Chandra61} requires that the specific angular momentum must be a
monotonically increasing function with respect to radius.  The gas
with the lowest $j$ progressively piles up in the center of DM
potential wells until \tcool~$<$~\tdyn~when it can catastrophically
cool and collapse.  Such low $j$ gas may originate in lower mass
progenitors because the gas resided in shallow potential wells
(i.e. low mass halos) that led to smaller turbulent and thermal
velocities.  We argue that this effect is intimately linked to the gas
acting to achieve virial equilibrium at all stages during the collapse
(see Paper I).  Furthermore, the system becomes unstable to turbulence
as the material segregates.  This onset of turbulence can be delayed
if viscosity is large enough so that Reynolds numbers are below the
order of $10^2$ or $10^3$.  However there are many modes of
instability if the Rayleigh criterion is not met, and even gas with a
low Reynolds number will eventually become fully turbulent on a
timescale that is chaotic, depending on the initial perturbation and
Reynolds number \citep{Shu92, Moehlis04}.  We note that a more
comprehensive approach would consider the Solberg-H\o iland criterion
\citep{Endal78} that generalizes this to include partial rotational
and pressure support in a disc.

3. \textit{Bar-like rotational instabilities}--- After sufficient
amounts of gas have migrated to small radii because of angular
momentum segregation, this gas increases its rotational velocity as it
conserves angular momentum.  Gas with similar angular momentum now
obtains some organized rotational velocity.  As the rotational energy
increases, some shells may become rotationally unstable ($T/\vert W
\vert \ge 0.14$) in a secular $m=2$ mode.  In the case of a collapsing
gas cloud, turbulent viscosity provides the dissipative force that
drives the secular instability.  The system then deforms into a
bar-like object, where the gas with large $j$ moves to larger radius
and gas with small $j$ can infall to even smaller radii.

\medskip

The combination of these three processes alleviates the ``angular
momentum problem'' of inside-out collapses.  Such a scenario of
angular momentum transport during a self-similar collapse may be
widely applicable in cosmological collapse problems.

\subsection{Secular Instability Cascade}

Our simulations follow the self-similar collapse of protogalactic
halos over 14 orders of magnitude in length.  We find that a cascade
of three (four) bar-like instabilities occur during the latter stages
of the collapse.  The ratios of mass enclosed in each successive
instability varies from 10 to 10,000 in our simulations.  As a
consequence of these instabilities, the collapse of the densest point
never halts because of rotational support.  Instead the gas becomes
rotationally unstable when it gains sufficient rotational energy.  The
lowest $j$ gas then falls to smaller radius and may become unstable
yet again.  This sequence could repeat itself several times.  In
addition, we find that rotational instabilities are possible without a
global disc as in simulation A.

This is the ``bars within bars'' scenario originally proposed to fuel
active galactic nuclei through dynamical rotational bar-like
instabilities \citep{Shlosman89, Shlosman90, Heller07}.  It was then
adapted for funneling enough gas into pre-galactic ($M \sim 10^5 \Ms$)
SMBHs by \citet{Begelman06}, in whose framework the angular momentum
of the disc, where the instability occurs, depends on the spin
parameter of the halo \citep[see also][]{Mo98}.  Thus the amount of
gas available for accretion onto the central SMBH also depends on the
spin parameter.  Dynamical instabilities require 45\% more rotational
energy to occur than secular ones.  In the framework of
\citeauthor{Begelman06}, only requiring secular instabilities may
result in a larger fraction of halos forming a pre-galactic SMBH
because of the log-normal distribution of spin parameters
(eq. \ref{eqn:lambda_prob}).  Nevertheless, we do not advocate our
simulations as evidence of pre-galactic SMBH formation because we have
neglected many important processes related to \hh~cooling and
primordial star formation that we detail briefly in the next section.

\subsection{Applicability}
\label{sec:applicability}

\subsubsection{Limitations of Current Approach}
\label{sec:limitation}

Our results depict the importance of turbulence, accretion, and the
hydrogen cooling in the initial collapse of these halos.  However we
are missing some essential processes, such as \hh~chemistry,
primordial and Population II stellar formation and feedback, SMBH
formation and feedback, and metal transport and cooling.  It was our
intention to study only the hydrogen and helium cooling case first and
gradually introduce other processes at a later time to investigate the
magnitude and characteristics of their effects, which we will present
in later papers.

Gas becomes optically thick to \lya~radiation above column densities
of $\sim$$10^{13}$ cm$^{-2}$, and \lya~radiation trapping becomes
important above a density of $\sim$$5 \times 10^8 \cubecm$
\citep{Oh02}.  We continue to use optically thin cooling rates above
this density.  Thus we overestimate the cooling within 0.03 pc.  As a
consequence, we do not suggest that these simulated objects ever form
in nature.  However this scenario poses an excellent numerical
experiment of turbulent collapse, which should be common in galaxy
formation, where turbulence is generated during virialization, and
star formation within turbulent molecular clouds.

\subsubsection{Desired Improvements}
\label{sec:improve}

Clearly local dwarf spheroidals contain stars with ages consistent
with formation at very high redshifts \citep{Ferrara00, Tolstoy02,
  Tolstoy03, Helmi06}.  To develop a model that desires to fit galaxy
luminosity functions down to the faintest observed galaxies one may
need a star formation and feedback model that follows molecular clouds
as small as one thousand solar masses in order to allow for the
dominant mode of star formation observed locally.  It should be
already technologically feasible with current cosmological
hydrodynamical models to simulate these galaxies one star at a time.

Correct initial conditions for early galaxy formation require prior
star and BH formation and feedback.  The typically adopted conditions
for phenomenological star formation are velocity convergence, a
critical overdensity, \tdyn~$>$ \tcool, and being Jeans unstable
\citep{Cen92}.  Phenomenological primordial star formation is possible
if we include two additional conditions as utilized in \citet{Abel07}.
First, the \hh~fraction must exceed 10$^{-3}$ \citep{Abel02a}, and
second, the metallicity of the gas must not exceed some ``critical
metallicity'' of 10$^{-3}$ -- 10$^{-6}$ of the solar value
\citep{Bromm01, Schneider06, Smith07, Jappsen07a, Jappsen07b}.  From
prior studies \citep[e.g.][]{Abel02a, Bromm03, OShea05, Greif06}, we
expect these stars to form to in halos that can support \hh~cooling
and ones embedded in relic \ion{H}{2} regions.  The Lyman-Werner
radiation from massive stars can dissociate \hh~from large distances
\citep{Dekel87, Haiman00}, suppress star formation in lower mass halos
\citep{Machacek01, Wise05}, and should be considered to accurately
model future star formation.

BH formation in the death of some primordial stars can also have a
profound effect on surrounding structure formation as it accretes
infalling matter during later mergers.  In principle, one should
include feedback of seed BHs from primordial stars with masses outside
of the range between 140 and 260 solar masses.  Also it is possible to
phenomenologically model SMBH formation in a similar manner as the
stellar case.  If the protogalactic collapse occurs faster than
stellar formation timescale of a massive star, a SMBH may form inside
this region.  Using the stellar formation conditions plus this
condition and allowing the particle to accrete \citep[i.e. sink
particles;][]{Bate95, Krumholz04}, protogalactic collapses can be
followed in cosmological hydrodynamic simulations \citep{Clark07}.
These sink particles should regulate the accretion with an appropriate
subgrid model.  Important processes include an appropriate accretion
rate (e.g. Eddington or Bondi-Hoyle), turbulence \citep{Krumholz06},
rotational support of the infalling gas, and a viscosity timescale for
accretion discs.

For small galaxies, radiative transfer effects can have a great impact
\citep[e.g.][]{Haehnelt95, Whalen04, Kitayama04, Alvarez06} and should
not be neglected.  Ionization front instabilities in these galaxies
create cometary small-scale structure and shadowing effects as a
result from the explicit treatment of three-dimensional radiation
hydrodynamics.  Stellar feedback can have both a positive and negative
impact on subsequent star formation.  Some examples of positive
feedback include enhanced \hh~formation in relic \ion{H}{2} regions
\citep{Ferrara98, OShea05, Johnson07} and dust and metal-line cooling
\citep{Glover03, Schneider06, Jappsen07a}.  Negative feedback may
occur from baryonic expulsion from host halos \citep{Whalen04,
  Kitayama04, Yoshida07, Abel07} and halo photoevaporation
\citep{Susa06, Whalen08}.

The promising approach of \citet{Gnedin01} has recently been
implemented and coupled with the AMR hydrodynamic code ART
\citep{Gnedin08}.  Also, the technique of adaptive ray tracing
\citep{Abel02b} has been implemented into \enzo~and used to study the
outflows and ionizing radiation from a primordial star \citep{Abel07}.
This method has also been independently implemented into \enzo~by
\citet{Razoumov06}.  Finally as used in many stellar formation
routines \citep{Cen92, Tassis03}, we hope to include thermal and
radiative feedback from Population II stars in future studies.

\section{CONCLUSIONS}

We have simulated the hydrodynamics and collapse of a protogalactic
gas cloud in two cosmology AMR realizations.  Our focus on the
hydrodynamics presents a basis for future studies that consider
stellar and BH feedback.  In the idealized case presented, we find a
central dense object forms on the order of 10$^5 \Ms$ and $r \lsim 5$
pc.  This central object is not rotationally supported and does not
fragment in our simulations.  However our results do not dismiss disc
formation in protogalaxies because rotationally supported disc
formation may begin after the initial central collapse.  Disc
formation may be sensitively affected by feedback from the central
object.

These simulations highlight the relevance of secular bar-like
instabilities in galaxy formation and turbulent collapses.  Similar
bar structures are witnessed in primordial star formation simulations.
As low angular momentum infalls, it gains rotational energy as it
conserves angular momentum.  This induces an $m$ = 2, bar-like
instability that transports angular momentum outwards, and the
self-similar collapse can proceed without becoming rotationally
supported and exhibits a density profile $\rho \propto r^{-12/5}$.
This process repeats itself as material infalls to small scales that
is indicative of the ``bars within bars'' scenario.  We see three and
four occurrences of embedded secular instabilities in the two
realizations studied here.

We also find that supersonic turbulence influences the collapse by
providing a channel for the gas to preferentially segregate according
to its specific angular momentum.  The low angular momentum material
sinks to the center and provides the material necessary for a central
collapse.  Here the possibilities of a central object include a direct
collapse into a SMBH \citep[e.g.][]{Bromm03}, a starburst
\citep[e.g.][]{Clark07}, or a combination of both
\citep[e.g.][]{Silk98}.  All of these cases are viable in the early
universe, and the occurrence of these cases depends on the merger
history, local abundances in the halo, and the existence of a seed BH.
Moreover, star formation should occur whether a central BH exists or
not.  Perhaps the frequency of these different protogalactic outcomes
may be traced with either 3D numerical simulations that consider star
and SMBH formation and feedback along with metal transport or Monte
Carlo merger trees that trace Pop III star formation, metallicities,
and BHs.  We will attempt the former approach in future studies to
investigate protogalactic formation in more realistic detail.

\acknowledgments 

This work was supported by NSF CAREER award AST-0239709 from the
National Science Foundation.  We thank Kristen Menou, Michael Norman,
Ralph Pudritz, Darren Reed, and Tom Theuns for helpful discussions.
We applaud Greg Bryan and Michael Norman for developing an incredible
code.  The clarity and presentation of this paper was greatly improved
by constructive comments from an anonymous referee.  We are grateful
for the continuous support from the computational team at SLAC.  We
performed these calculations on 16 processors of a SGI Altix 3700 Bx2
at KIPAC at Stanford University.

{} 

\end{document}